\documentclass[twocolumn,tighten,twocolappendix]{aastex631}
\usepackage{CJK}
\graphicspath{{./}{figures/}}
\begin{document}

\title{A Luminous Red Supergiant and Dusty Long-period Variable Progenitor for SN\,2023ixf}

\author[0000-0001-5754-4007]{Jacob E.\ Jencson}
\affil{Department of Physics and Astronomy, Johns Hopkins University, 3400 North Charles Street, Baltimore, MD 21218, USA}
\affil{Space Telescope Science Institute, 3700 San Martin Drive, Baltimore, MD 21218, USA}

\author[0000-0002-0744-0047]{Jeniveve Pearson}
\affil{Steward Observatory, University of Arizona, 933 North Cherry Avenue, Tucson, AZ 85721-0065, USA}

\author[0000-0003-4666-4606]{Emma R.\ Beasor}\altaffiliation{Bok Fellow}
\affil{Steward Observatory, University of Arizona, 933 North Cherry Avenue, Tucson, AZ 85721-0065, USA}

\author[0000-0003-0778-0321]{Ryan M.\ Lau}
\affil{NSF's NOIRLab, 950 North Cherry Avenue, Tucson, AZ 85719, USA}

\author[0000-0003-0123-0062]{Jennifer E.\ Andrews}
\affil{Gemini Observatory/NSF's NOIRLab, 670 N. A'ohoku Place, Hilo, HI, 96720, USA}

\author[0000-0002-4924-444X]{K.\ Azalee Bostroem}\altaffiliation{LSSTC Catalyst Fellow}
\affil{Steward Observatory, University of Arizona, 933 North Cherry Avenue, Tucson, AZ 85721-0065, USA}

\author[0000-0002-7937-6371]{Yize Dong \begin{CJK*}{UTF8}{gbsn}(董一泽)\end{CJK*}\!\!}
\affil{Department of Physics and Astronomy, University of California, 1 Shields Avenue, Davis, CA 95616-5270, USA}

\author[0000-0003-0209-674X]{Michael Engesser}
\affil{Space Telescope Science Institute, 3700 San Martin Drive, Baltimore, MD 21218, USA}

\author[0000-0001-6395-6702]{Sebastian Gomez}
\affil{Space Telescope Science Institute, 3700 San Martin Drive, Baltimore, MD 21218, USA}

\author[0000-0002-5063-0751]{Muryel Guolo}
\affil{Department of Physics and Astronomy, Johns Hopkins University, 3400 North Charles Street, Baltimore, MD 21218, USA}

\author[0000-0003-2744-4755]{Emily Hoang}
\affil{Department of Physics and Astronomy, University of California, 1 Shields Avenue, Davis, CA 95616-5270, USA}

\author[0000-0002-0832-2974]{Griffin Hosseinzadeh}
\affil{Steward Observatory, University of Arizona, 933 North Cherry Avenue, Tucson, AZ 85721-0065, USA}

\author[0000-0001-8738-6011]{Saurabh W.\ Jha}
\affil{Department of Physics and Astronomy, Rutgers, the State University of New Jersey, 136 Frelinghuysen Road, Piscataway, NJ 08854-8019, USA}

\author[0000-0003-2758-159X]{Viraj Karambelkar}
\affiliation{Division of Physics, Mathematics and Astronomy, California Institute of Technology, Pasadena, CA 91125, USA}

\author[0000-0002-5619-4938]{Mansi M.\ Kasliwal}
\affiliation{Division of Physics, Mathematics and Astronomy, California Institute of Technology, Pasadena, CA 91125, USA}

\author[0000-0001-9589-3793]{Michael Lundquist}
\affil{W.\ M.\ Keck Observatory, 65-1120 Mamalahoa Highway, Kamuela, HI 96743-8431, USA}

\author[0000-0002-7015-3446]{Nicolas E.\ Meza Retamal}
\affil{Department of Physics and Astronomy, University of California, 1 Shields Avenue, Davis, CA 95616-5270, USA}

\author[0000-0002-4410-5387]{Armin Rest}
\affil{Department of Physics and Astronomy, Johns Hopkins University, 3400 North Charles Street, Baltimore, MD 21218, USA}
\affil{Space Telescope Science Institute, 3700 San Martin Drive, Baltimore, MD 21218, USA}

\author[0000-0003-4102-380X]{David J.\ Sand}
\affil{Steward Observatory, University of Arizona, 933 North Cherry Avenue, Tucson, AZ 85721-0065, USA}

\author[0000-0002-9301-5302]{Melissa Shahbandeh}
\affil{Department of Physics and Astronomy, Johns Hopkins University, 3400 North Charles Street, Baltimore, MD 21218, USA}
\affil{Space Telescope Science Institute, 3700 San Martin Drive, Baltimore, MD 21218, USA}

\author[0000-0002-4022-1874]{Manisha Shrestha}
\affil{Steward Observatory, University of Arizona, 933 North Cherry Avenue, Tucson, AZ 85721-0065, USA}

\author[0000-0001-5510-2424]{Nathan Smith}
\affil{Steward Observatory, University of Arizona, 933 North Cherry Avenue, Tucson, AZ 85721-0065, USA}

\author[0000-0002-1468-9668]{Jay Strader}
\affil{Center for Data Intensive and Time Domain Astronomy, Department of Physics and Astronomy, Michigan State University, East Lansing, MI 48824, USA}

\author[0000-0001-8818-0795]{Stefano Valenti}
\affil{Department of Physics and Astronomy, University of California, 1 Shields Avenue, Davis, CA 95616-5270, USA}

\author[0000-0001-5233-6989]{Qinan Wang}
\affil{Department of Physics and Astronomy, Johns Hopkins University, 3400 North Charles Street, Baltimore, MD 21218, USA}

\author[0000-0002-0632-8897]{Yossef Zenati}\altaffiliation{ISEF \& De Gunzburg Fellowship}
\affil{Department of Physics and Astronomy, Johns Hopkins University, 3400 North Charles Street, Baltimore, MD 21218, USA}



\begin{abstract}
We analyze pre-explosion near- and mid-infrared (IR) imaging of the site of SN\,2023ixf in the nearby spiral galaxy M101 and characterize the candidate progenitor star. The star displays compelling evidence of variability with a %\edit1{
possible %}
period of $\approx$1000~days and an amplitude of $\Delta m \approx 0.6$~mag in extensive monitoring with the Spitzer Space Telescope since 2004, 
%\N{likely} 
likely indicative of radial pulsations. %\N{(give amplitude...)} 
Variability consistent with this period is also seen in the near-IR $J$ and $K_{s}$ bands between 2010 and 2023, up to just 10~days before the explosion. Beyond the periodic variability, we do not find evidence for any IR-bright pre-supernova outbursts in this time period. The IR brightness ($M_{K_s} = -10.7$~mag) and color ($J-K_{s} = 1.6$~mag) of the star suggest a luminous and dusty red supergiant. Modeling of the phase-averaged spectral energy distribution (SED) yields constraints on the stellar temperature ($T_{\mathrm{eff}} = 3500_{-1400}^{+800}$~K) and luminosity ($\log L/L_{\odot} = 5.1\pm0.2$). This places the candidate among the most luminous Type II supernova progenitors with direct imaging constraints, with the caveat that many of these rely only on optical measurements. %\N{(although this higher mass is partly due to the available multiwavelength SED... say this if true(?). in other words, many HST-only masses may be underestimates.)} 
Comparison with stellar evolution models gives an initial mass of $M_{\mathrm{init}} = 17\pm4~M_{\odot}$. We estimate the pre-supernova mass-loss rate of the star between 3 and 19~yr before explosion %\N{(at 3-11 yr pre explosion - this is important, because it may change 1 yr before...)} 
from the SED modeling at $\dot M \approx 3\times10^{-5}$ to 
 $3\times10^{-4}~M_{\odot}$~yr$^{-1}$ for an assumed wind velocity of $v_w = 10$~km~s$^{-1}$, perhaps pointing to enhanced mass loss in a pulsation-driven wind.
\end{abstract}

\keywords{Supernovae(1375) --- Massive stars(732) --- Stellar mass loss(1613) --- Evolved stars(481) --- Circumstellar dust(236)}


\section{Introduction}\label{sec:intro}
The direct identification of core-collapse (CC) supernova (SN) progenitors in archival, pre-explosion imaging provides a vital test of our understanding of stellar evolution. To date, detections of $\sim$25 progenitor candidates have been reported (see, e.g., \citealp{smartt2015,vandyk2017} for recent reviews and references therein), with several now confirmed to have disappeared in late-time imaging (e.g., \citealp{vandyk2023a} and references therein). A great success of this decades-long effort has been the confirmation that the progenitors of the most common class of CC SNe, the hydrogen-rich Type II-plateau and II-linear SNe (SNe~II-P and II-L), are massive ($>$8~$M_{\odot}$) red supergiants (RSGs), in excellent agreement with predictions from stellar evolutionary theory. 

Still, unexpected questions have emerged as the sample of SN~II-P and II-L progenitors has grown. Specifically, the sample seems to consist of RSGs of only modest initial masses $\lesssim$18~$M_{\odot}$ \citep[e.g.,][]{smartt2009,smartt2015} even though observed populations of RSGs extend to $\gtrsim$25~$M_{\odot}$ \citep{humphreys1979,davies2018,mcdonald2022}. The apparent lack of higher-mass progenitors, dubbed the ``RSG problem,'' remains controversial. Numerous solutions have been proposed, including the direct collapse of higher-mass progenitors to black holes, effects related to the uncertain environmental or circumstellar extinction, the difficulties of connecting limited observations to uncertain stellar models, and questions regarding the statistical validity of the upper-mass limit itself \citep[e.g.,][]{davies2007,kochanek2008,smith2011,walmswell2012,davies2018,davies2020,davies2020a,kochanek2020}.

An important component of tying CC SNe to their massive progenitors is an understanding of stellar mass loss during the final evolutionary phases. Material surrounding the star as it approaches CC can dramatically affect both the appearance of the progenitor and the observable properties of the SN \citep[e.g.,][]{kochanek2012,smith2014,davies2022}. Growing evidence from SN observations, namely early ``flash'' spectroscopy \citep{gal-yam2014,yaron2017,bruch2021,tartaglia2021}, numerical light-curve modeling \citep{morozova2017,morozova2018,subrayan2023}, and instances of observed pre-SN activity (e.g., \citealp{kilpatrick2018,jacobson-galan2022,matsumoto2022} though see also evidence for progenitor stability in, e.g., \citealt{johnson2018,tinyanont2019}), all point to dense circumstellar material (CSM) around SN~II progenitors. Several mechanisms for enhanced mass loss have been proposed to explain the presence of this material. These include nuclear burning instabilities, enhanced pulsation-driven winds, wave-driven mass loss, and neutrino-driven mass loss 
%\N{(add citation to Smith \& Arnett 2014)}
\citep{heger1997,yoon2010,arnett2011,quataert2012,shiode2013,moriya2014a,shiode2014,smith2014a,smith2014,woosley2015,fuller2017,wu2021}. 

Here, we analyze extensive pre-explosion near- and mid-infrared (IR) imaging and characterize a candidate progenitor star of SN\,2023ixf. Discovered on 2023 May 19.73 (all dates UT) by \citet{itagaki2023} and located in M101 ($D = 6.85\pm0.15$~Mpc; $\mu = 29.18$~mag; \citealp{riess2022}), SN\,2023ixf is one of the nearest SNe~II of the last decade. As an exceptionally well-studied nearby galaxy, M101 has a rich archival data set, allowing us to study the photometric evolution of the progenitor in the final years before explosion---a vitally important phase for which it is rarely possible to obtain direct constraints. Intensive early monitoring of SN\,2023ixf already shows evidence of dense CSM \citep{berger2023,bostroem2023,grefenstette2023,jacobson-galan2023,smith2023,teja2023,yamanaka2023}.
%\N{and Smith et al. in prep.  actually it is submitted to arxiv}). %In Section~\ref{sec:data}, we describe 
The Milky Way extinction toward SN~2023ixf is $E(B-V)_\mathrm{MW} = 0.0077$~mag \citep{schlafly2011}, and we adopt a foreground, host galaxy extinction of $E(B-V)_\mathrm{host} = 0.03$ \citep{smith2023}. The value for the host extinction is consistent with that reported by \citet[][$E(B-V)_\mathrm{host} = 0.031 \pm 0.006$~mag]{lundquist2023}. We correct for these using the extinction law of \citet{fitzpatrick1999} with $R_V = 3.1$.

\section{Data and Observations} \label{sec:data}

\subsection{Spitzer Imaging}
\begin{figure*}
\centering
\includegraphics[width=\textwidth]{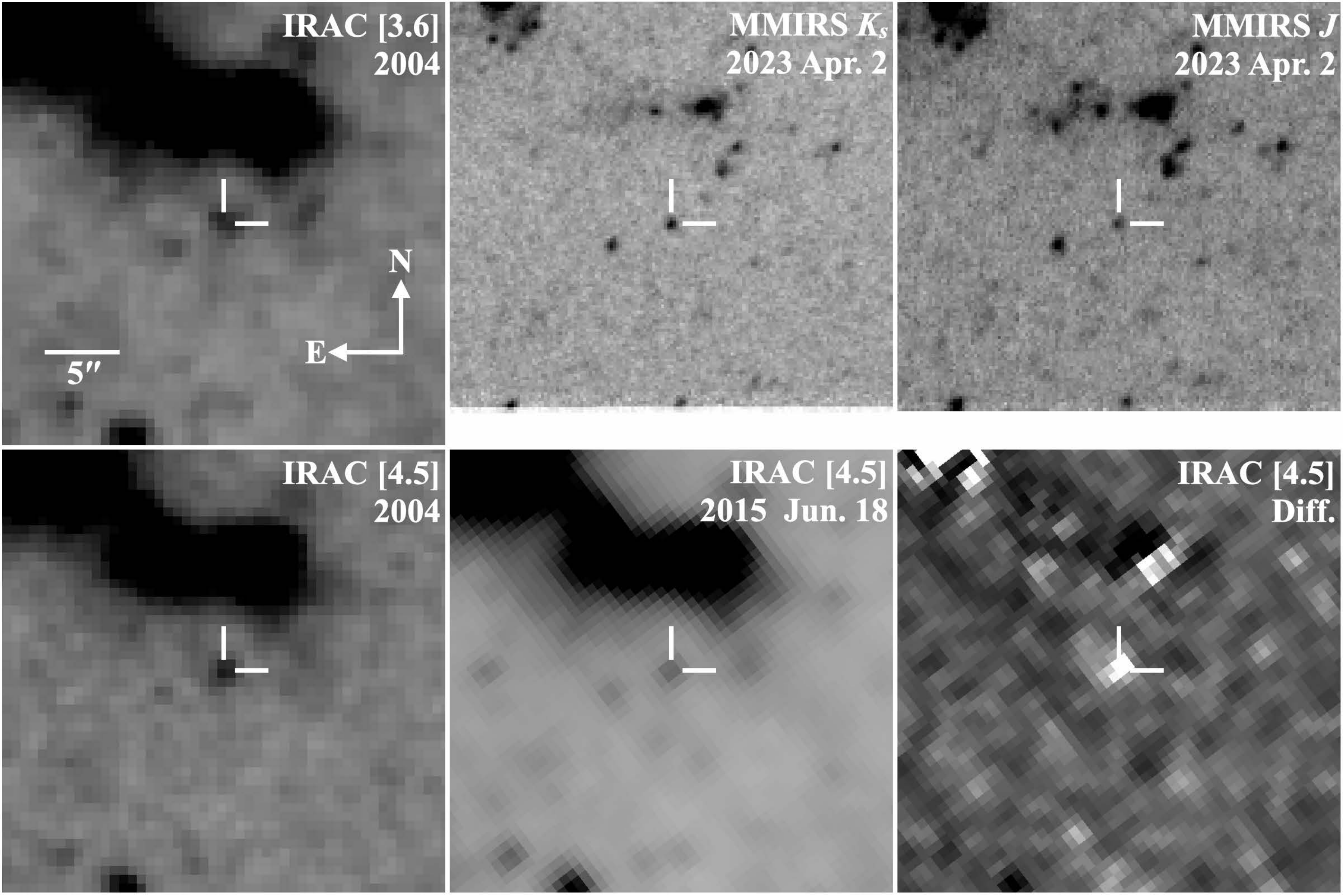}
\caption{\label{fig:images}
IR pre-explosion imaging of the site of SN\,2023ixf. The source identified as the progenitor candidate is indicated by the white crosshairs at the center of each panel. In the left column, we show the Spitzer/IRAC [3.6] (top) and [4.5] (bottom) Super Mosaics. In the bottom center and right panels, respectively, we show the [4.5] image from 2015 June 18 and its corresponding difference image, where the Super Mosaic was used as the template for subtraction. The negative (white) flux at the position in the difference image indicates the source was fainter in 2015 than in 2004. The top center and right panels are the MMIRS $J$- and $K$-band images from 2023 April 2. The orientation and scale of each image are the same as indicated in the upper left panel.
}
\end{figure*}

The location of SN\,2023ixf (R.A., decl.: $14^{\mathrm{h}}03^{\mathrm{m}}38\fs56, +54\degr18\arcmin41\farcs9$, J2000.0) was imaged by the Spitzer Space Telescope \citep{werner2004,gehrz2007} during the cold mission in all four channels (3.6, 4.5, 5.8, and 8.0~$\mu$m; [3.6], [4.5], [5.8], and [8.0], respectively) of the Infrared Array Camera (IRAC; \citealp{fazio2004}) on 2004 March 8.32 (PI: G.\ Rieke, PID 60). These images were stacked into Super Mosaics\footnote{Super Mosaics are available as Spitzer Enhanced Imaging Products through the NASA/IPAC Infrared Science Archive \citep{SEIP_DOI}.} along with images taken on 2007 December 31, but as only the 2004 images cover the SN position, we consider that date to be the effective time of these observations for our photometric measurements. As part of a previous study on IR variability in nearby galaxies, PSF-fitting photometry source catalogs were made for M101 using the Super Mosaics in all four bands \citep{karambelkar2019}. An empirical model of the PSF for each Mosaic was made using the \texttt{DAOPHOT/ALLSTAR} package \citep{stetson1987}, with corrections for the finite radius of the model PSF (following the method of \citealp{khan2017}, and see \citealp{karambelkar2019} for more details). As shown in Figure~\ref{fig:images}, a source is clearly visible at the location at [3.6] and [4.5] and is detected in the PSF catalogs at $[3.6] = 17.5 \pm 0.1$ and $[4.5] = 16.75 \pm 0.04$\,mag.\footnote{All photometry is reported on the Vega system.} This source was previously reported by \citet{szalai2023}, and its position was confirmed to be consistent with that of the SN by \citep{kilpatrick2023}. We consider this star as a strong candidate progenitor of SN\,2023ixf. 

There is no clear point-like source visible in the longer wavelength channels, and no detections were recovered at the position in the respective photometry catalogs. As the detection limit is likely dominated by the significant background emission, we adopt upper limits as the measured surface brightness at the position integrated over the size of the PSF plus 2 times the image rms at a position of blank sky. This yields limiting magnitudes of $[5.8] > 14.1$ and $[8.0] > 11.8$\,mag, using the Vega system zero-magnitude fluxes defined in the IRAC Instrument Handbook \citep{IRACHandbook_DOI}.%\footnote{\url{http://irsa.ipac.caltech.edu/data/SPITZER/docs/irac/iracinstrumenthandbook/}}

The SN position was then imaged numerous times at [3.6] and [4.5] since 2012 during the warm mission by multiple programs (PI: P.\ Garnavich, PID 80126; PI: M.\ Kasliwal, PIDs 80196, 90240), including with frequent monitoring of M101 between 2014 and the end of 2019 by the SPitzer Infrared Intensive Transients Survey (SPIRITS; PI: M.\ Kasliwal, PIDs 10136, 11063, 13053, 14089). The post-basic calibrated data level images were downloaded from the Spitzer Heritage Archive \citep{SHA_DOI}
%\footnote{\url{https://sha.ipac.caltech.edu/applications/Spitzer/SHA/}}
and Spitzer Early Release Data Service\footnote{\url{https://irsa.ipac.caltech.edu/data/SPITZER/Early_Release/}} and processed through an automated image-subtraction pipeline (for survey and pipeline details, see \citealp{kasliwal2017,jencson2019}). The Super Mosaics were used as template images for the subtractions. An example difference image is shown in Figure~\ref{fig:images}, demonstrating the variability of the source. We performed aperture photometry on our difference images adopting the appropriate aperture corrections from the IRAC Handbook and following the method for a robust estimate of the photometric uncertainties as described in \citet{jencson2020}. We sum our difference flux measurements with the reference PSF-photometry measurement on the Super Mosaics, again using the handbook-defined values to convert to Vega-system magnitudes, to produce our final light curves of the source shown in Figure~\ref{fig:lcs}. 

\subsection{Ground-based, Near-IR Imaging}
We obtained imaging of M101 in the near-IR $J$ and $K_s$ bands with the MMT and Magellan Infrared Spectrograph (MMIRS, $0\farcs2$ pixels; \citealp{mcleod2012}) on the 6.5~m MMT Observatory telescope on Mt.\ Hopkins in Arizona %at the Smithsonian's Fred Lawrence Whipple Observatory 
at multiple epochs in 2021--2023. These images were taken as part of an ongoing program to monitor the unusual variable star and failed SN candidate M101-OC1 reported by \citet{neustadt2021} and serendipitously covered the location of SN\,2023ixf prior to the explosion. The last images were taken on 2023 May 9.37, just 10.36 days before the discovery of the SN. 

Each observation consisted of dithered sequences alternating between the target position on M101 and an offset blank-sky field every few minutes to allow for accurate subtraction of the bright near-IR sky background. We reduced the images using a custom pipeline\footnote{Adapted from the MMIRS imaging pipeline developed by K.\ Paterson, available here: \url{https://github.com/CIERA-Transients/Imaging_pipelines}} that performs standard dark-current subtraction, flat-fielding, sky background estimation and subtraction, astrometric alignments, and final stacking of the individual exposures. 

We also downloaded $J$- and $K_\mathrm{cont}$-band imaging with the Near-Infrared Imager\footnote{\url{http://www.gemini.edu/instrumentation/niri}} (NIRI) on the 8~m Gemini-N Telescope on Maunakea from the Gemini Observatory Archive. The images were taken with the f/6 camera ($0\farcs117$ pixels) on 2010 April 18 (PI: Bosch; PID GN-2010A-Q-27). We reduced all the images using \texttt{DRAGONS} \citep{labrie2023}, %\footnote{\url{https://dragons.readthedocs.io/en/v3.1.0/index.html}}
a Python-based platform for reducing Gemini data, and following the procedures for extended sources outlined in the NIRI imaging-reduction tutorial.\footnote{\url{https://dragons.readthedocs.io/projects/niriimg-drtutorial/en/stable/}} 

As shown in Figure~\ref{fig:images}, a bright, point-like source is visible at the position of the SN in both $J$ and $K_s$. The star is consistent with the location of the Spitzer source, and again, the same star was identified in the NIRI imaging by \cite{kilpatrick2023}. The field-of-view (FOV) of the MMIRS imager ($6\farcm9 \times 6\farcm9$) is sufficient to calibrate the photometric zero-points using aperture photometry of relatively isolated stars in images with cataloged $JHK_\mathrm{s}$-band magnitudes in the Two Micron All Sky Survey (2MASS; \citealp{skrutskie2006}). We then derived a model of the effective (e)PSF for each image by fitting bright, isolated stars using the \texttt{EPSFBuilder} tool of the \texttt{photutils} package in \texttt{Astropy}. We performed PSF-fitting photometry at the location of the candidate progenitor as well as for a set of approximately 60 stars spread across the images with varying degrees of crowding and galaxy-background emission. We include a low-order, two-dimensional polynomial in the fit to account for the spatially varying background for each star, taking care to avoid overfitting the data. We adopt the rms error of the fit residuals, scaled by a factor of the square root of the reduced $\chi^2$ (typically $\gtrsim 1$) for the fit, as the nominal statistical uncertainty per pixel, and multiply by the effective footprint, or a number of ``noise pixels,'' of the ePSF\footnote{A derivation of this quantity is provided by F.\ Masci here: \url{http://web.ipac.caltech.edu/staff/fmasci/home/mystats/noisepix_specs.pdf}} to obtain an estimate of the statistical uncertainty for each flux measurement. We used the set of 2MASS calibration stars to derive aperture corrections ($\lesssim0.1$\, mag in all three filters) to place the PSF-fitting magnitudes on the scale of the image photometric zeropoints. We adopt the statistical flux uncertainty, summed in quadrature with the rms error of the stars used in estimations of the zero-point and ePSF aperture correction, as the total uncertainty in our final magnitudes. Owing to the limited number of isolated 2MASS stars, even with the large FOV of MMIRS, the zero-point rms (typically $\approx$0.1\,mag) dominates the error budget. 

\begin{figure*}
\centering
\includegraphics[width=\textwidth]{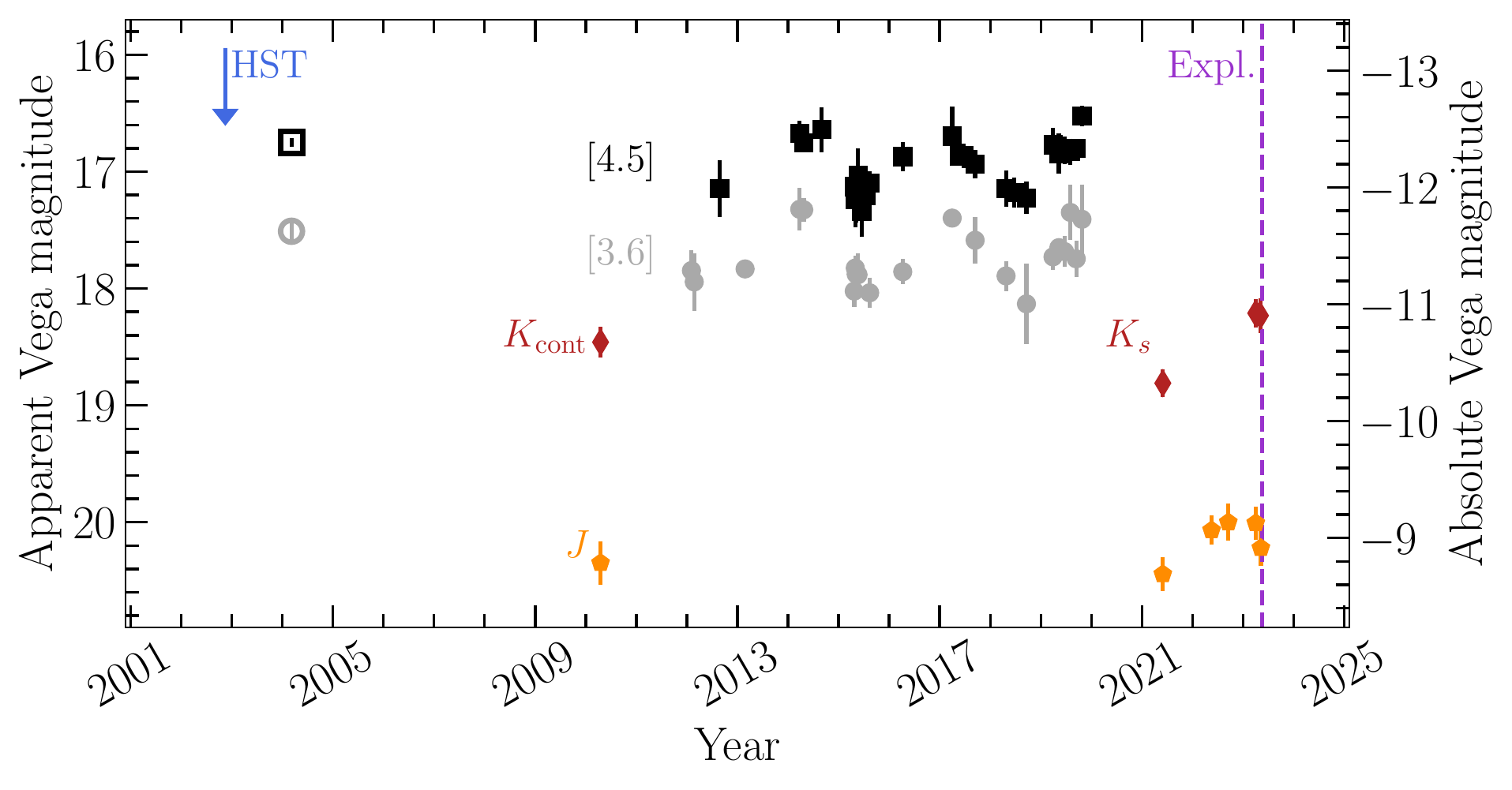}
\caption{\label{fig:lcs}
Mid- and near-IR pre-explosion light curves of the SN\,2023ixf progenitor candidate. Mid-IR [3.6] and [4.5] Spitzer light curves are based on image subtraction (filled gray circles and black squares, respectively) relative to PSF photometry on the 2004 Super Mosaics (open symbols). The $J$- and $K_s/K_{\mathrm{cont}}$-band light curves (orange pentagons and red diamonds, respectively) from ground-based NIRI and MMIRS imaging extend up to just 10.35 days before the SN discovery, indicated by the purple dashed line. The epoch of archival 2002 HST imaging reported by \citet{soraisam2023}, \citet{pledger2023}, and \citet{kilpatrick2023} is indicated by the blue downward arrow. (The data used to create this figure are available in the published article.) 
}
\end{figure*}

The FOVs of the NIRI ($\approx$2$\arcmin\times$2$\arcmin$) images are smaller, and there were not enough isolated 2MASS stars in the field to do a direct calibration. We instead cross-calibrated our PSF photometry of stars in these images, performed in the same manner as described above, to a set of $\approx$10 common stars with the corresponding MMIRS image in the same filter (the $K_{\mathrm{cont}}$ NIRI image is calibrated to MMIRS $K_s$). We then adopted the statistical uncertainty from the PSF fitting (as above), summed in quadrature with the zero-point uncertainty (from the standard deviation of the individual stars used in the cross-calibration) as our measurement uncertainty. All of our near-IR photometry are shown in Figure~\ref{fig:lcs}. 

\section{Pre-explosion Light Curves and Variability}\label{sec:lc_var}
As shown in Figure~\ref{fig:lcs}, the IR pre-explosion light curves extend back almost two decades prior to the explosion of SN\,2023ixf. The Spitzer light curves display clear variability with an apparent periodicity of $\approx$3~yr and full amplitudes of 0.6~mag at [3.6] and 0.45~mag at [4.5] (as also recently reported by \citealp{kilpatrick2023}). Between 2021 and 2023 May (less than two weeks before the SN), the near-IR $J$ and $K_s$-band light curves also appear to brighten by 0.5 and 0.6~mag, respectively. It is not immediately clear whether this is part of the same periodic variability observed by Spitzer or indicative of a small outburst in the final few years before the explosion. %\N{(If it is an outburst, it has a pretty small amplitude... this is actually pretty important....)}

To test for periodicity, we simultaneously fit the [3.6] and [4.5] light curves using the Lomb--Scargle method \citep{lomb1976,scargle1982} implemented in \texttt{Astropy}, restricting our search to sinusoidal signals. The resulting periodogram peaks at a best-fitting period of $P=1119.4$~days with Lomb--Scargle power of 0.75. The peak is split, with a nearby secondary period of $P=967.6$~days at a lower score of 0.67. No other peak in the power spectrum has a score higher than 0.35. The sinusoidal fit (reduced $\chi^2 = 1.02$) provides a significantly better fit to the Spitzer data over a null hypothesis of a constant flux in each band (reduced $\chi^2 = 3.3$), %\edit1{
supporting the possibility of periodic variability in the light curve. %}.
%indicating that there is good evidence for periodic variability in the light curve.
We adopt the periods at half the maximum power as upper and lower bounds for the uncertainty %\edit1{
on the possible period, %},  
giving $P=1119.4_{-233.3}^{+132.4}$~days. %\edit1{
This result is consistent with that reported by \citet{soraisam2023a} from an independent analysis of the Spitzer data.%}

\begin{figure}
\centering
\includegraphics[width=0.5\textwidth]{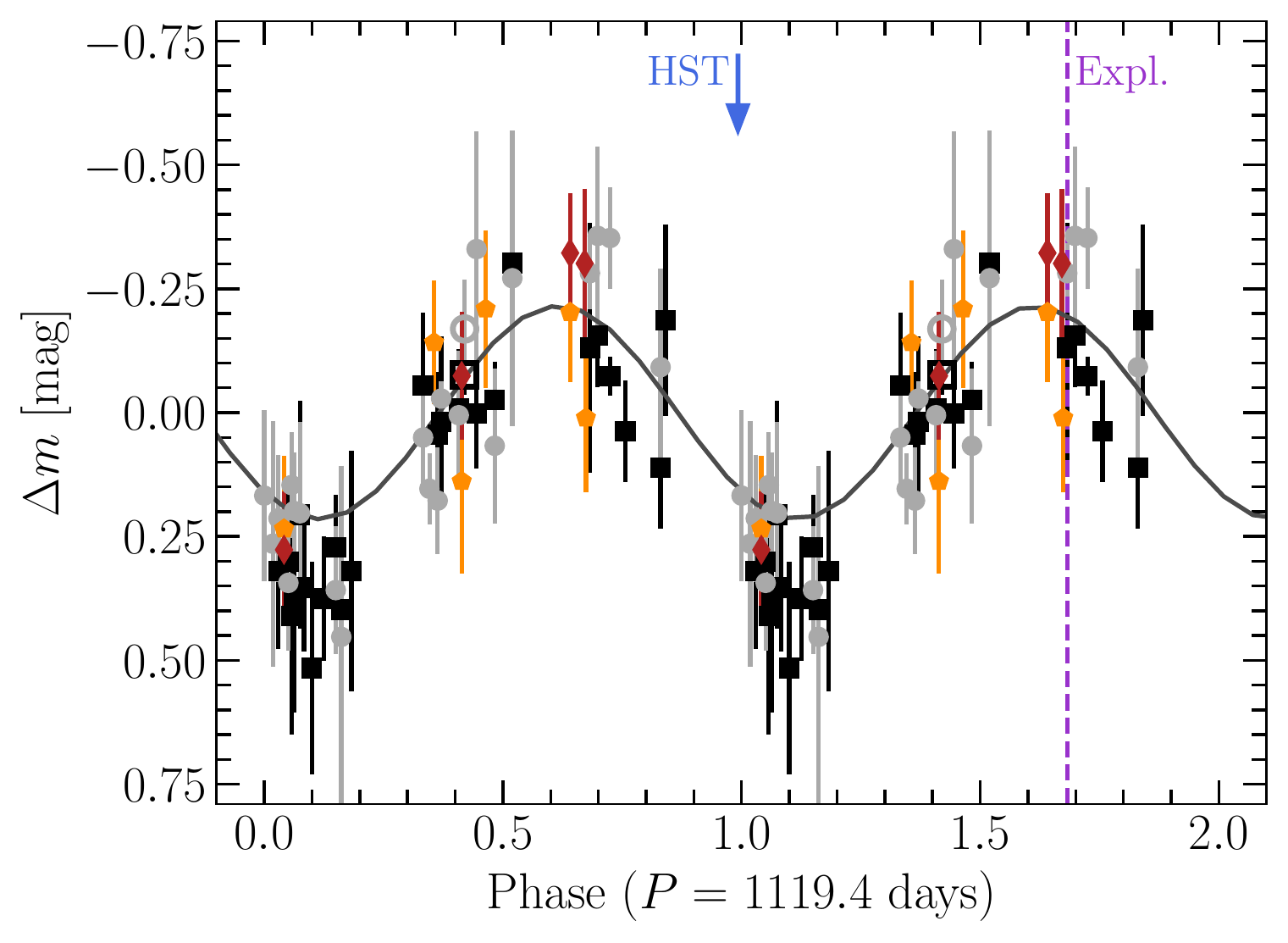}
\caption{\label{fig:phased_lcs}
IR light curves (symbols are the same as in Figure~\ref{fig:lcs}) folded to a best-fitting period of $P=1119.4$~days from a joint Lomb--Scargle analysis of the [3.6] and [4.5] data. A phase-weighted average magnitude has been subtracted out for each band. The ground-based near-IR data agree remarkably well with the pulsation cycle derived from the Spitzer measurements, without any additional fine-tuning of the parameters. The inferred phases of the explosion epoch and archival HST observations are indicated by the purple dashed line and blue downward arrow, as in Figure~\ref{fig:lcs}. 
}
\end{figure}

In Figure~\ref{fig:phased_lcs}, we fold all of the IR light curves with the same best-fitting period, where the phase-weighted average magnitude has been subtracted out for each band. The ground-based near-IR data agree remarkably well with the periodic cycle derived for the Spitzer light curves without any additional tuning of the parameters. This provides strong evidence that the brightening seen just prior to the explosion in the $J$ and $K_s$-bands is part of a normal pulsation cycle of the star. We see no clear evidence for any outbursts or eruptive variability up to just 10~days before the SN. %\N{(this is crazy.  i mean, it is cool, but it makes my life more difficult.   where the heck did the CSM come from then?  we should emphasize that the lack of a bright eruption is actually an important and very constraining result...Ne/O burning, etc....maybe you will later.)}

RSGs commonly exhibit periodic light-curve variations attributed to radial pulsations \citep{stothers1969,stothers1971,guo2002}, with more luminous RSGs typically exhibiting longer periods and larger optical amplitudes \citep{kiss2006,yang2011,yang2012,soraisam2018}. In the IR, \citet{karambelkar2019} extended the known period--luminosity correlations for long-period variable stars \citep[e.g.,][]{riebel2015,goldman2017} to higher luminosities ($M_{[4.5]} < -12$~mag) and longer periods ($>$1000~days). They postulated that the brightest of these sources may be dusty RSGs or the so-called super-AGB stars from massive ($\approx$8--12~$M_{\odot}$; \citealp{siess2007,doherty2015,doherty2017}) progenitors. Super-AGBs may be expected to exhibit the reddest IR colors ($[3.6]-[4.5] \gtrsim 1$~mag), longest periods ($\gtrsim$1500~days) and largest amplitudes $\Delta m\gtrsim1.5$. Given its high luminosity ($M_{[4.5]} \approx -12.3$), the relatively more modest color ($[3.6]-[4.5] = 0.8\pm0.1$~mag), 1000~day period, and amplitudes ($\approx$0.6~mag) of the SN\,2023ixf progenitor candidate are likely more consistent with a dusty RSG (see also Section~\ref{sec:colors} below). 

%Massive RSGs ($\gtrsim17~M_{\odot}$) have been predicted to undergo enhanced pulsational-driven winds in their final evolutionary stages, resulting in a ``superwind'' phase leading up to their explosion as CC SNe \citep{heger1997,yoon2010,förster2018}. The long $\approx$1000~day observed period, as well as the evidence for dusty CSM (see further discussion in Section~\ref{sec:mlr}), may constitute the strongest direct evidence for this phenomenon in an SN progenitor candidate seen to date. 
%\N{(you don't need to discuss this here, or at all if you don't want to, but ... the fact that the narrow lines go away in a few days means that the dense CSM is very confined radially, and this limits the heavy mass loss to just a couple years pre-SN, given the observed CSM velocities... it would be very hard to tune the yoon \& cantiello pulsations to only work for the last couple years...especially if we want the same mechanism to apply to many of these SNe...)}

\subsection{IR Photometric Classification and Bolometric Correction}\label{sec:colors}
Based on our periodicity analysis above, we compute foreground-extinction-corrected, phase-weighted mean magnitudes in each band of $J=20.18\pm0.20$\,mag, $K_s = 18.52 \pm 0.19$\,mag, $[3.6] = 17.67 \pm 0.18$\,mag, and $[4.5] = 16.82 \pm 0.16$\,mag. To account for the uncertainty in the period and the variability amplitude in each band, we incorporate a 15\% uncertainty summed in quadrature with that of the best individual measurement (the 2023 April 2 measurements from MMIRS and the 2004 Spitzer Super Mosaics). At $M_{K_s} = -10.7$\,mag, the star is firmly above the tip of the red giant branch (TRGB) and, moreover, is brighter than nearly all asymptotic giant branch (AGB) stars identified in nearby galaxies including the Large and Small Magellanic Clouds (L/SMCs), M31, and M33 \citep[see, e.g.,][and references therein]{cioni2006,boyer2011,massey2021}. Its near-IR color of $J-K_\mathrm{s} = 1.6\pm0.28$\,mag is redder than the range typically used to discriminate RSGs from luminous AGBs. As noted in \citet{boyer2011}, however, it is essentially impossible to distinguish a dusty RSG, which will be very red in $J-K_s$, from an AGB star, with IR photometry alone. \citet{massey2021} argue that stars brighter than $M_{K_s} = -10$\,mag are likely RSGs even at redder colors, as they are more luminous than expected for the brightest AGB stars. Based on this, and because of its likely association with the Type II SN\,2023ixf, we find that the progenitor candidate is most likely an RSG that suffers additional reddening from a dense molecular wind or circumstellar dust. 

The $K_s$ band is useful as a luminosity indicator for RSGs, both because the effects of extinction are reduced compared to optical bands and because the bolometric correction, $BC_K$, is found empirically to be constant across early-to-late M-type supergiants in Milky Way and LMC star clusters \citep{davies2018}. Assuming an M-type spectrum ($T_{\rm eff} \lesssim 3700$\,K) and adopting their value of $BC_K = m_{bol} - m_{K} = 3.0$\,mag, we obtain bolometric luminosities of $\log (L/L_{\odot}) \approx 5.0$. Given the red colors of the star described above, the true bolometric correction may be smaller if there is excess circumstellar extinction. Still, this value can likely be viewed as a robust lower limit on the luminosity. We discuss the possible locations of the star in a Hertzsprung--Russell diagram (HRD) based on modeling of the spectral energy distribution (SED) below in Section~\ref{sec:SED}.

\section{SED Modeling}\label{sec:SED}
\begin{figure*}
\centering
\includegraphics[width=\textwidth]{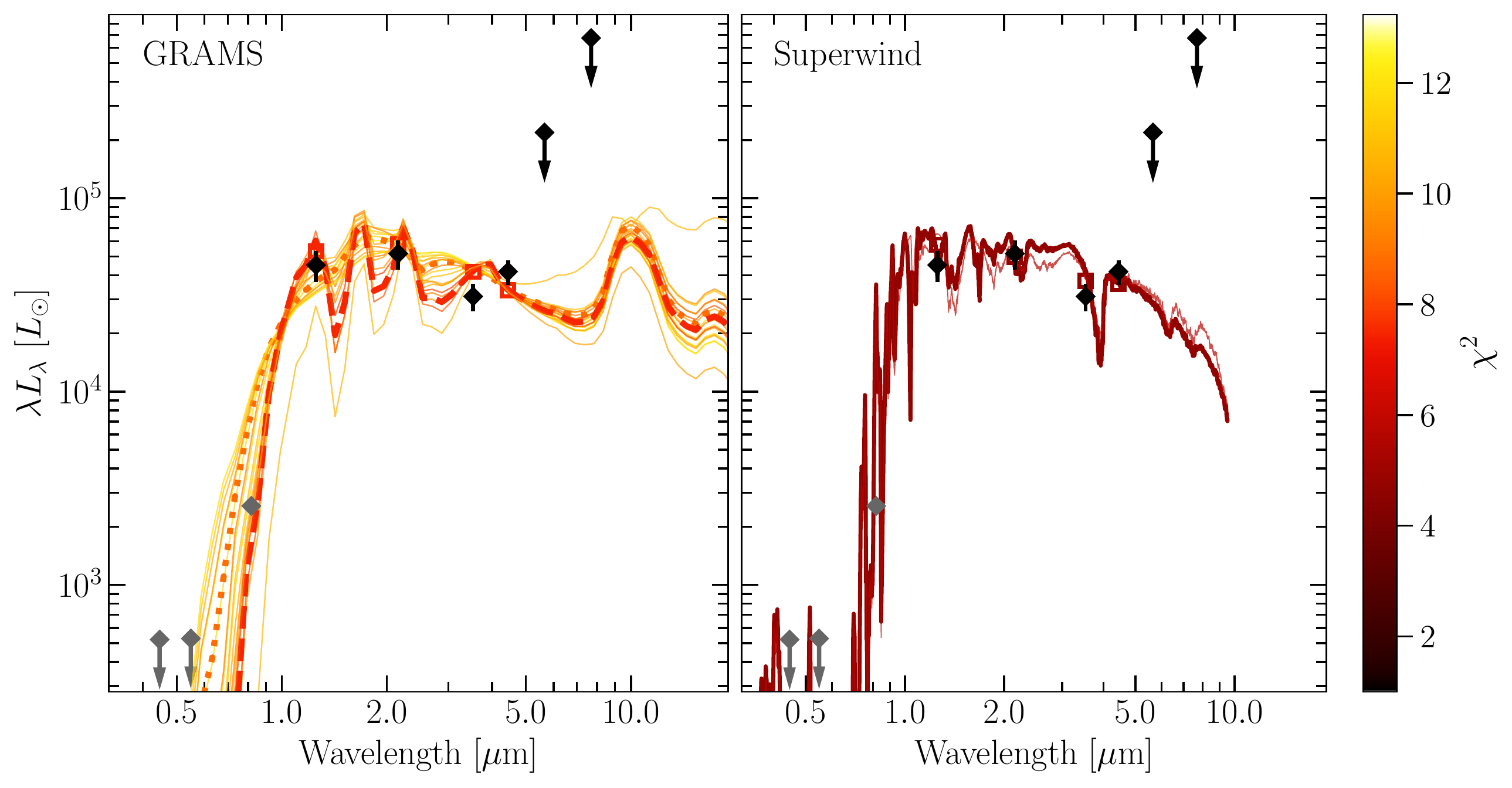}
\caption{\label{fig:SED_GRAMS}
The SED of the SN2023\,ixf progenitor candidate is shown in both panels, including phase-weighted average flux measurements from Spitzer and MMIRS (black diamonds) and the HST measurements reported by \citet[][gray diamonds]{kilpatrick2023}. Downward arrows indicate upper limits. In the left panel, the 20 best-fitting GRAMS models to the IR data points are shown as curves, where the color indicates $\chi^2$ for the fit as indicated by the color bar. The best-fitting single model ($T_{\mathrm{eff}} = 2300$~K, $\log L/L_{\odot} = 5.07$) is indicated as the thick dashed curve, and the corresponding synthetic photometry in the four IR bands included in the fitting are shown as the red open square symbols. A warmer model ($T_{\mathrm{eff}} = 3500$~K, $\log L/L_{\odot} = 5.12$) at a secondary, relative minimum in the $\chi^2$ distribution is shown as the thick dotted curve. In the right panel, we show two superwind models from \citet{davies2022}, with the same mapping of color to $\chi^2$. 
}
\end{figure*}

In Figure~\ref{fig:SED_GRAMS}, we construct an SED of the progenitor candidate from the phase-averaged magnitude measurements in the ground-based near-IR and Spitzer bands. The photometric magnitudes were converted to luminosities, $\lambda L_{\lambda}$, using the filter transmission curves compiled by the Spanish Virtual Observatory (SVO) Filter Profile Service\footnote{Documentation for the SVO Filter Profile Service is available at \url{http://ivoa.net/documents/Notes/SVOFPSDAL/index.html}} to compute zero-point fluxes and effective wavelengths for each filter. We also show Hubble Space Telescope (HST) measurements for the progenitor candidate reported by \citet{kilpatrick2023}, which, based on our best-fitting period, may have been timed near the bottom of the pulsation cycle (See Figure~\ref{fig:phased_lcs}). Given the significant uncertainty in the amplitude of any optical variability, we do not include these points in the fitting procedure described below but note that luminosity estimates based primarily on the HST data may be underestimates. %\N{(understood, however, you have a period solution, so you can at least say something about what phase of the variability the HST detection is at.  i.e. do we expect that the HST mag should be a little fainter or brighter than the average, based on the period?  i know the visual amplitude may be different, but saying this at least helps put it in perspective...that may impact how we interpret differences in Mzams estimates...)}

%\begin{figure}
%\centering
%\includegraphics[width=0.5\textwidth]{figures/SN2023ixf_prog_SED_GRAMS.pdf}
%\caption{\label{fig:SED_GRAMS}
%SED of the SN2023\,ixf progenitor candidate, including phase-weighted average flux measurements from Spitzer and MMIRS (black diamonds) and the HST measurments reported by \citet[][gray diamonds]{pledger2023}, where downward arrows indicate upper limits. The 20 best-fitting GRAMS models to the IR data points are shown as curves, where the color indicates $\chi^2$ for the fit as indicated by the color bar. The best-fitting single model ($T_{\mathrm{eff}} = 2300$~K, $\log L/L_{\odot} = 5.07$) is indicated as the thick dashed curve and the corresponding synthetic photometry in the four IR bands included in the fitting are shown as the red open square symbols. A warmer model ($T_{\mathrm{eff}} = 3500$~K, $\log L/L_{\odot} = 5.12$) at a secondary, relative minimum in the $\chi^2$ distribution is shown as the thick dotted curve. 
%}
%\end{figure}

The SED is very red, peaking in the near-IR between the $J$ and $K_s$ bands. To estimate the physical parameters of the star, we fit the SEDs with the Grid of Red supergiant and Asymptotic Giant Branch ModelS (GRAMS; \citealp{sargent2011,srinivasan2011}) using a similar procedure to that described in \citet{jencson2022}. This suite of radiative transfer models consists of a base grid of 1225 spectra from spherically symmetric shells of varying amounts of silicate dust \citep[][appropriate for RSGs]{ossenkopf1992} around stars of constant mass-loss rates computed using the dust radiative transfer code \texttt{2-Dust} \citep{ueta2003}. The grid employs input PHOENIX model photospheres \citep{kucinskas2005,kucinskas2006} for 1\,$M_{\odot}$ stars (model spectra can be scaled for more luminous and massive, i.e., supergiant, stars) with effective temperatures, $T_{\mathrm{eff}}$, between 2100 and 4700\,K, and at a fixed subsolar metallicity\footnote{The metallicity of the input stellar models in GRAMS was chosen to be similar to the LMC, while a value closer to solar may be more appropriate for the environment in a large spiral galaxy of SN\,2023ixf. We do not expect this to significantly affect the shape of broad-band SEDs, however, and we believe our estimates of stellar parameters ($L$, $T_{\mathrm{eff}}$) will not depend strongly on this choice \citep[see discusions in, e.g.,][]{beasor2016,vandyk2019,jencson2022}.} $\log(Z/Z_{\odot}) = -0.5$ and a fixed surface gravity $\log g = -0.5$. The amount of circumstellar dust is characterized in terms of the optical depth at 1~$\mu$m, $\tau_1$, from which a dust mass-loss rate, $\dot M_{\mathrm{d}}$ is inferred assuming a wind speed of $v_w = 10$\,km\,s$^{-1}$. The inner radius of the dust shell, $R_{\mathrm{in}}$, takes values of 3, 7, 11, and 15 times the stellar radius, $R_*$. 

For each model in the grid, we compute the scale factor that minimizes the value of $\chi^2$ between the IR data points and synthetic photometry derived from the model for each filter. The 20 best models with the lowest $\chi^2$ values are compared to the data in Figure~\ref{fig:SED_GRAMS}. These models span $T_{\mathrm{eff}} = 2100$--4300~K and $\log L/L_{\odot} = 5.01$--5.22, while the single best-fitting model (minimum $\chi^2 = 7.5$) has $T_{\mathrm{eff}} = 2300$~K and $\log L/L_{\odot} = 5.08$. As a function of $T_{\mathrm{eff}}$, the $\chi^2$ distribution has a secondary local minimum ($\chi^2 = 8.7$) at 3500~K and a corresponding luminosity of $\log L/L_{\odot} = 5.12$.

\subsection{Position in the HRD and Initial Mass}\label{sec:HRD}
\begin{figure}
\centering
\includegraphics[width=0.5\textwidth]{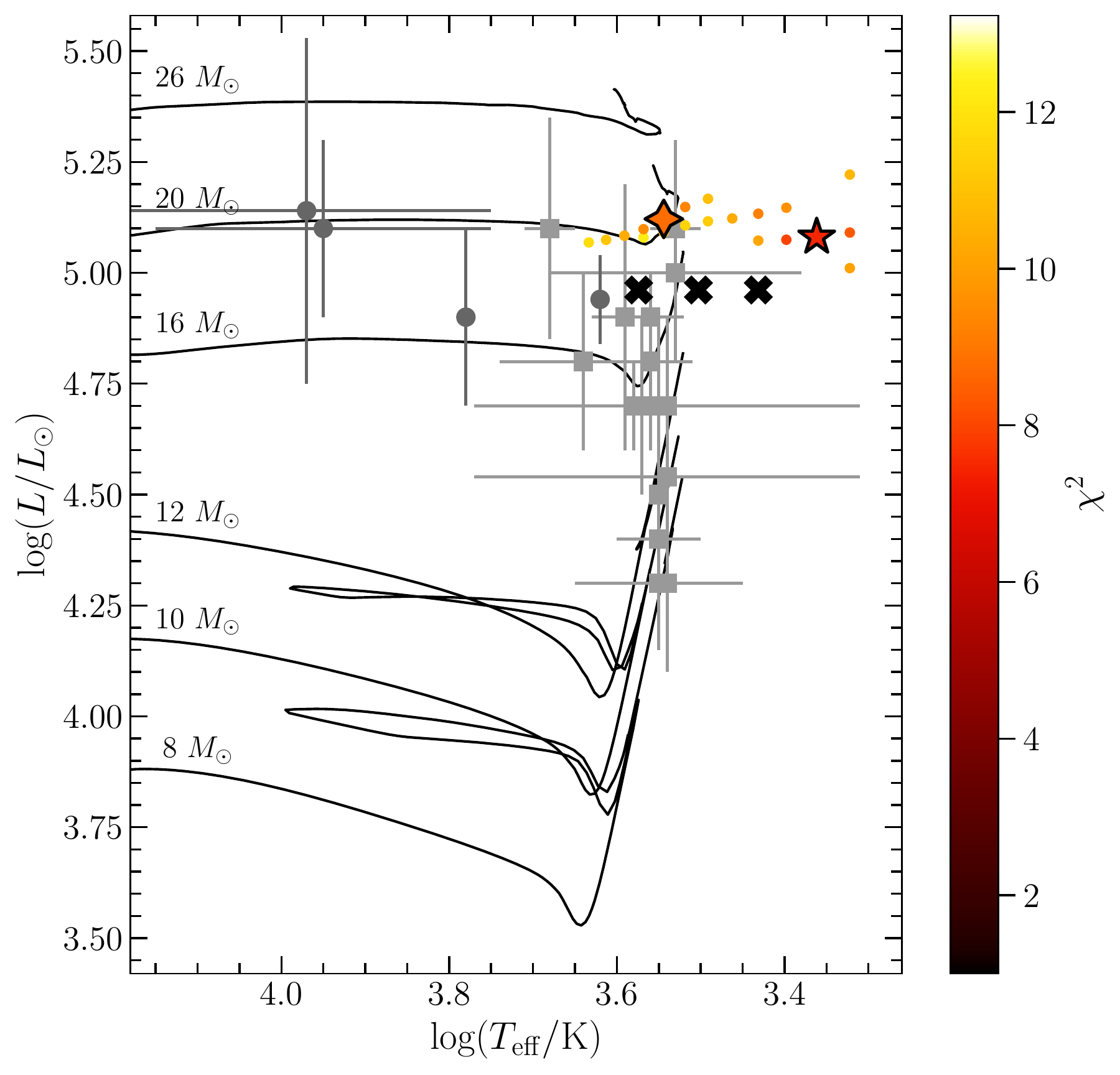}
\caption{\label{fig:HRD} HRD showing possible locations of the SN\,2023ixf progenitor candidate. The result for the single best-fitting model from our SED analysis is shown as the large red star symbol, while the range of the 20 best models are shown by the small circles. The orange four-pointed star represents a secondary, local minimum in the $\chi^2$ distribution at a more typical temperature for RSGs. The color of each point represents the $\chi^2$ for the fit, as indicated by the color bar. The ``$\times$'' symbols show the locations of the star assuming early, mid, and late M-type spectra and applying the $K$-band bolometric corrections of \citet{davies2018} to the average $K_s$-band measurement. We show stellar evolutionary tracks from MIST (nonrotating, solar metallicity) for a set of massive stars in the range $M = 8$--26\,$M_{\odot}$ as black curves for comparison. We also show the collection of directly detected SN progenitors of Types II (light gray squares) and IIb (dark gray circles; see text in Section~\ref{sec:HRD}). %\N{(maybe not in the caption, but somewhere in the text we should emphasize that your estimates for 2023ixf come from a full SED, whereas many of the HST progenitors are just 1 or 2 optical bands.... they could easily be systematic underestimates.  this should be kept in mind when comparing your estimate to the others.  don't you want to put charlie's logL=4.74 on here...? )} 
}
\end{figure}

In Figure~\ref{fig:HRD}, we place the progenitor candidate in the HRD based on the results of our SED modeling described above in Section~\ref{sec:SED}. Several models, including the single best-fitting model at $T_{\mathrm{eff}} = 2300$~K, are colder ($T_{\mathrm{eff}} < 3000$~K) than expected for end-stage RSGs, appearing far to the right of the terminal points of the stellar tracks derived from the Mesa Isochrones and Stellar Tracks models (MIST; \citealp[][nonrotating, solar metallicity]{choi2016,choi2017}). This may be an effect of extended CSM mimicking the appearance of a cooler star, i.e., from strong molecular opacity from enhanced winds \citep[e.g.,][]{davies2021}, additional extinction and excess IR emission from dust \citep[e.g.,][]{{scicluna2015,massey2006,haubois2019}}, or a combination of both. We discuss these possibilities and the inferred mass-loss rates in more detail below in Section~\ref{sec:mlr}. 

Considering this, we adopt the best-fitting warmer model ($T_{\mathrm{eff}} > 3000$~K) with  $T_{\mathrm{eff}} = 3500$~K and $\log L/L_{\odot} = 5.12$ as our preferred model though the luminosity is well constrained regardless of the temperature. As shown in Figure~\ref{fig:HRD}, the progenitor candidate of SN\,2023ixf is among the most luminous RSG progenitors of a Type II SN with direct-detection constraints \citep[][]{smartt2015,vandyk2017,kilpatrick2017,kochanek2017,kilpatrick2018,oneill2019,rui2019,vandyk2019,sollerman2021,vandyk2023}. We emphasize though that many of these previous estimates were derived primarily using only a few (usually HST) optical bands and may systematically underestimate the luminosities of the stars without IR constraints on the SED. 

Our inferred luminosity for the SN\,2023ixf progenitor candidate is a factor of $\approx$2 higher than that recently reported by \cite{kilpatrick2023}. We suspect this difference is attributable to two main sources: (1) modest discrepancies in the NIRI $K_{\mathrm{cont}}$ and [4.5] magnitudes, coupled with our use of phase-averaged measurements to construct the SED, resulting in fluxes that are $\approx$30--40\% higher in those bands and (2) our use of O-rich SED models, which contain significant flux in the silicate ``bump'' near 10~$\mu$m that is absent from the graphitic models used by \cite{kilpatrick2023}. %\N{(what the... why would they use graphite for RSG CSM dust?  hmmmm...)} 
%\edit1{
In contrast, \citet{soraisam2023a} infer an even higher luminosity ($\log L/L_{\odot} = 5.2$ to 5.5) by applying the period--luminosity relation for RSGs of \citet{soraisam2018}.%}

Comparing our range of luminosities to the end points of MIST evolutionary tracks (accounting for an additional $\approx$2.3\% uncertainty in the distance from \citealp{riess2022}), we infer an initial mass in the range $M_{\mathrm{init}} = 17\pm4~M_{\odot}$. 

\subsection{Constraints on Pre-SN Mass Loss}\label{sec:mlr}
All but one of the 20 best models have $\tau_1 = 2.2$ and resulting dust mass-loss rates between $\dot M_{\mathrm{d}} = 1.6 \times 10^{-7}$ and $1.4 \times 10^{-6}$~$M_{\odot}$~yr$^{-1}$. A single model with $\tau_1 = 4.45$ is the reddest model shown in Figure~\ref{fig:SED_GRAMS}, which significantly underpredicts the $J$-band flux. Assuming a gas-to-dust ratio of 200 (appropriate for Milky Way RSGs; \citealp{vanloon2005}), this corresponds to mass-loss rates in the range $\dot M = 3\times10^{-5}$ to $3\times10^{-4} \left(\frac{v_w}{10~\mathrm{km~s}^{-1}}\right)$~$M_{\odot}$~yr$^{-1}$ ($1.6\times10^{-3} \left(\frac{v_w}{10~\mathrm{km~s}^{-1}}\right)$~$M_{\odot}$~yr$^{-1}$ for the $\tau_1 = 4.45$ model). 

%\N{(can you also quote a total CSM within some radius?  because the velocities are assumed, of course, and the total CSM mass you need for CSM interaction is a little more robust than Mdot, so it would be interesting to quote what you get from dust.     ---- in any case, there should also be a caveat here that the derived dust masses and Mdots are probably lower limits and can be significant underestimates if the CSM is asymmetric or clumpy.  if you have a dense disk, for example, or optically thicj clumnps, then the dust can be self shielding and there might be a lot of cooler dust mass that you don't see in the IRAC 1 and 2 bands because it is not hot enough or optically thick to the star's radiation... the longer wavelength limits are not super constraining...)}

These values are elevated compared to modern estimates of mass-loss rates for normal RSGs in this luminosity range; the mass and luminosity dependent prescription of \citet{beasor2020}, for example, gives $\dot M \sim 10^{-6}$~$M_{\odot}$~yr$^{-1}$ for our preferred stellar parameters ($M_{\mathrm{init}} = 17$~$M_{\odot}$, $\log L/L_{\odot} = 5.1$). They are more in line with those of dusty, OH-IR stars (massive AGB and RSG stars exhibiting circumstellar maser emission and IR excesses) in the LMC and Galactic center/bulge \citep{goldman2017}. The red $J-K_{s}$ color (see Section~\ref{sec:colors}) points to a dusty RSG with significant CSM, so the enhanced mass-loss rates inferred here are perhaps unsurprising. 

%\edit1{
Our estimates are largely consistent with those derived from early X-ray observations ($3\times10^{-4}$~$M_\odot$~yr$^{-1}$ for $v_w=50$~km~s$^{-1}$; \citealp{grefenstette2023}) but somewhat %} 
lower than those for confined CSM ($\lesssim$10$^{15}$~cm) from early observations of ``flash'' features in the spectra of SN\,2013ixf ($\dot M \sim 10^{-3}$--10$^{-2}$~$M_\odot$~yr$^{-1}$% \edit1{
for $v_w=50$~km~s$^{-1}$; %}; 
\citealp{bostroem2023,jacobson-galan2023}). The exact values depend strongly on multiple assumptions (e.g., gas-to-dust ratio, in addition to $v_w$). %Notably, the assumed value of $v_w = 10$~km~s$^{-1}$ is lower than that used in the analysis of \citet[][25~km~s$^{-1}$]{beasor2020} or \citet[][50~km~s$^{-1}$]{jacobson-galan2023}, and assuming a higher value would increase our estimates closer to those derived from the SN spectra. 
%\edit1{
We note as well that %} 
a larger mass of self-shielding cold dust, emitting beyond the shorter wavelength [3.6] and [4.5] IRAC channels, could also be hidden if the CSM is highly aspherical or clumpy, leading to underestimates of the mass-loss rate in the SED fitting.  
%\N{(FYI - echelle spectra indicate that the CSM expansion speed is more like 115 km/s at least... eek! but that might not have been the expansion speed a decade pre-SN.)}

Our analysis of the IR variability of the progenitor candidate over the last $\approx$13~yr (Section~\ref{sec:lc_var}) indicates that the star was %\edit1{
likely %}
undergoing steady, long-period pulsations. Importantly, there is no indication of any outbursts between 3 and 11~yr pre-explosion in Spitzer monitoring, nor are there any large changes in the near-IR fluxes or colors up to only 10.4~days before the SN. Optical imaging also places stringent limits on any long-lasting outbursts up to $\approx$400~days pre-explosion \citep{neustadt2023}. This apparent stability is inconsistent with predictions from early observations of the SN for increased activity of the progenitor over the last $\sim$3~yr \citep[][]{jacobson-galan2023}. Instead, our findings point to a steady but enhanced wind that develops over the final $\gtrsim$decade of the star's life. %Intriguingly, the upper end of our derived intial mass range (17--20~$M_{\odot}$) and long observed period ($\approx$1000~days) are broadly consistent with the predictions for end-stage, pulsation-driven ``superwinds'' in massive RSGs \citep{yoon2010}. 
For our assumed velocity, the extent of the CSM from such a wind lasting at least 13~yr would be $R_{\mathrm{CSM}} > 4\times10^{14}$~cm. 
%\N{(sounds like the language here is favoring the yoon \& cantiello pulsations.  however, some caution:  the Y\&C runaway pulsations kick in when the L/M ratio gets high (i.e. eddingotn limit) because of a lot of previous mass loss. they predict that the star will end its life with a tiny H envelope (like 0.5 Msun in the H envelope for their 20 Msun model).  this is ruled out if you want the SN to have a plateau.... in any case, it is a prediction that the SN light curve should indicate a tiny H envelope mass if this is valid.)}

We explore a ``superwind'' scenario \citep[as described in][]{forster2018} using the models presented in \citet{davies2022}. These models use a 3800~K MARCS stellar atmosphere \citep{gustafsson2008} and attach a wind following a $\rho \propto r^{-2}$ density profile. The velocity of the wind is taken to be a $\beta$ law as a function of radius (see Eq.~2 in \citealp{davies2022}). The chosen wind parameters are taken from \citet{forster2018} and assume a superwind where $\dot M = 10^{-3}~M_{\odot}$~yr$^{-1}$, $v_w = 10$~km~s$^{-1}$ and $\beta = 3$. The wind is propagated out to distances of $r_w$ = 3, 3.5, 5, 10 and 20~$R_*$,  where 20~$R_*$ roughly corresponds to 260~yr after the onset of the superwind. 

Unlike an outburst, a superwind can take hundreds of years before it begins to substantially affect the observed spectrum (see Fig.~2 in \citealp{davies2022}). In Figure~\ref{fig:SED_GRAMS}, we find that the red IR colors of the progenitor candidate are best matched by the models with $r_w = 10$ and 20~$R_*$ ($\chi^2 = 4.7$ and 5.7, respectively), while the other models give significantly worse fits ($\chi^2 \gtrsim 30$). This would imply the superwind was launched $\sim$200~yr prior to SN.  We note, however, that these models are dust free. The inclusion of dust, in combination with enhanced molecular opacity in the wind, may be able to produce redder spectra similar to that of the progenitor candidate without invoking these longer timescales. 

%\textit{Need to think about what this means a bit, compare to mass-loss estimates from the flash-papers, etc}.   \N{(i always say this, but... if the CSM is asymmetric (which it is), then you can have more mass in the CSM than would be indicated by column density along your line of sight (either TiO strength or dust extinction or whatever).)}

%I think I will mention comparing to early observations as part of the conclusions. 
%Early observations of the SN itself also indicate dense CSM in the immediate vicinity (e.g., K.~A.~Bostroem et al.\ 2023, in prep.; G.\ Hosseinzadeh et al.\ 2023, in prep.). 

%Uncertainties -> longer wavelength constraints would help, higher v_wind would increase estimates!
%Not surprising, red colors... also red spitzer color, could mention CO from CSM?

%mention models of outbursts or superwinds (e.g., davies 2022). Constrain CSM extent for wind. 
%SN also shows evidence of CSM (e.g., griffin and azalee in prep.)
%maybe this goes in the conclusions too

% 

\section{Summary and Conclusions}
We have identified a candidate progenitor of SN\,2023ixf as a bright IR source in archival Spitzer/IRAC and serendipitous ground-based near-IR imaging of M101 with MMT/MMIRS and Gemini-N/NIRI. In Spitzer, the star displays %clear 
evidence of a long $\approx$1000~day period since 2012, %\edit1{
likely %} 
indicative of radial pulsations. Variations seen in the near-IR $J$ and $K_{s}$ between 2010 and 2023---extending just 10~days before the SN discovery---are fully consistent with the Spitzer-derived pulsation period. There is no evidence for dramatic brightening due to eruptive, pre-SN outbursts, in tension with predictions from early SN observations \citep[e.g.,][]{jacobson-galan2023} and with those for instabilities on the timescales of the final nuclear burning stages \citep[e.g.][]{woosley2002}. The IR colors of the star are consistent with a luminous, highly evolved, and dusty RSG. Modeling of the phase-averaged SED of the star yields constraints on the stellar temperature ($T_{\mathrm{eff}} = 3500_{-1400}^{+800}$~K) and luminosity ($\log L/L_{\odot} = 5.1\pm0.2$), placing the candidate among the most luminous Type II SN progenitors with direct imaging constraints.  Comparison with stellar evolution models indicates an initial mass of $M_{\mathrm{init}} = 17\pm4~M_{\odot}$. We estimate the pre-SN mass-loss rate of the star from the SED modeling at $\dot M \approx 3\times10^{-5}$ to $3\times10^{-4} \left(\frac{v_w}{10~\mathrm{km~s}^{-1}}\right)$~$M_{\odot}$~yr$^{-1}$.  

%The inferred high initial mass, long pulsation period, and enhanced pre-SN mass-loss rate are all broadly consistent 
%\N{(see my comments above about why this is problematric...)} 
%with expectations for a pulsation-driven superwind \citep[e.g.,][]{heger1997,yoon2010}, the clearest direct evidence for this phenomenon in an RSG SN progenitor candidate known to date. 
%\N{(again, if the SN ends up being a Type II-P, then this is ruled out...)} 
Given the inferred high initial mass and long pulsation period, detailed comparisons with late-stage RSG pulsation models (see, e.g., recent work on Betelgeuse by \citealp{saio2023}) are a promising avenue to further constrain the fundamental stellar properties and inform their connection to enhanced pre-SN mass loss. Furthermore, as one of the nearest and brightest Type II SNe of the last decade, SN\,2013ixf will be exceptionally well observed. Already, early multi-wavelength data sets indicate clear signatures of interaction of the SN shock wave with dense, nearby CSM (\citealp{berger2023,bostroem2023,grefenstette2023,hosseinzadeh2023,jacobson-galan2023,smith2023,teja2023,yamanaka2023}; E.\ Zimmerman et al., in prep.). This will enable a rare chance to compare constraints on the surrounding CSM from SN observations directly to those from archival observations of the progenitor star and to recent theoretical predictions for the effects of enhanced CSM on observable properties of Type II SNe \citep[e.g.,][]{goldberg2020}. Continued monitoring to late times will be vital to both confirm the disappearance of the candidate progenitor and constrain any late-time interaction signatures that will extend our understanding of the full pre-SN mass-loss history %\edit1{
\citep[as in, e.g.,][]{rizzosmith2023}. %}. 
Altogether, we expect SN\,2023ixf to be a keystone object for interpreting early interaction signatures in CC SNe and connecting observations of their progenitors to the theory of massive star evolution and end-stage mass loss. 

\section{Acknowledgements}
%\vspace{0.5cm}
We thank B.\ Davies for sharing the superwind models and for illuminating discussions. We thank S.\ Points for advice regarding some of the near-IR photometry. We appreciate the efforts of the observing support staff of the MMT and for their help in planning and obtaining observations presented in this work. We also thank the anonymous referee for helpful comments. 

Time-domain research by the University of Arizona team and D.J.S.\ is supported by NSF grants AST-1821987, 1813466, 1908972, \& 2108032, and by the Heising-Simons Foundation under grant \#20201864. S.V.\ and the UC Davis time-domain research team acknowledge support by NSF grant AST-2008108. J.E.A.\ is supported by the international Gemini Observatory, a program of NSF's NOIRLab, which is managed by the Association of Universities for Research in Astronomy (AURA) under a cooperative agreement with the National Science Foundation, on behalf of the Gemini partnership of Argentina, Brazil, Canada, Chile, the Republic of Korea, and the United States of America. J.S. acknowledges support from the Packard Foundation. This publication was made possible through the support of an LSSTC Catalyst Fellowship to K.A.B., funded through Grant 62192 from the John Templeton Foundation to LSST Corporation. The opinions expressed in this publication are those of the authors and do not necessarily reflect the views of LSSTC or the John Templeton Foundation.

This work is based in part on archival data obtained with the Spitzer Space Telescope, which is operated by the Jet Propulsion Laboratory, California Institute of Technology, under a contract with NASA. Observations reported here were obtained at the MMT Observatory, a joint facility of the University of Arizona and the Smithsonian Institution. Based on observations obtained at the Gemini Observatory (Programs GN-2010A-Q-27), which is operated by the Association of Universities for Research in Astronomy, Inc., under a cooperative agreement with the NSF on behalf of the Gemini partnership: the National Science Foundation (United States), National Research Council (Canada), CONICYT (Chile), Ministerio de Ciencia, Tecnolog\'{i}a e Innovaci\'{o}n Productiva (Argentina), Minist\'{e}rio da Ci\^{e}ncia, Tecnologia e Inova\c{c}\~{a}o (Brazil), and Korea Astronomy and Space Science Institute (Republic of Korea). This research made use of Photutils, an Astropy package for detection and photometry of astronomical sources \citep{bradley2021}. 

\facilities{Gemini:Gillett (NIRI), MMT (MMIRS), Spitzer (IRAC)}

\software{\texttt{DAOPHOT/ALLSTAR} \citep{stetson1987},
\texttt{DRAGONS} \citep{labrie2023}, 
\texttt{Astropy} (\url{https://www.astropy.org/}; \citealp{astropycollaboration2013,astropycollaboration2018,astropycollaboration2022}), \texttt{photutils} \citep{bradley2021},
\texttt{2-Dust} \citep{ueta2003}} 

\bibliography{references_ascii}{}
\bibliographystyle{aasjournal}



\end{document}